\documentstyle[prl,aps,psfig]{revtex}
\begin{document}
\twocolumn[\hsize\textwidth\columnwidth\hsize\csname
@twocolumnfalse\endcsname
\title{
Expectation bubbles in a spin model of markets: 
\\
Intermittency from frustration across scales 
}
\author{Stefan Bornholdt}
\address{Institut f\"{u}r Theoretische Physik,
Universit\"at Kiel, Leibnizstrasse 15, D-24098 Kiel, 
Germany, bornholdt@theo-physik.uni-kiel.de
}
\maketitle
\begin{abstract}
A simple spin model is studied, motivated by 
the dynamics of traders in a market where expectation 
bubbles and crashes occur. The dynamics is governed by 
interactions which are frustrated across different scales: 
While ferromagnetic couplings connect each spin to 
its local neighborhood, an additional coupling 
relates each spin to the global magnetization. 
This new coupling is allowed to be anti-ferromagnetic. 
The resulting frustration causes a metastable dynamics 
with intermittency and phases of chaotic dynamics.  
The model reproduces main observations of real economic 
markets as power-law distributed returns and clustered 
volatility. 
\medskip \\ 
To be published in Int.\ J.\ Mod.\ Phys.\ C 12 (2001), No 5. 
\bigskip 
\end{abstract}
]

Market dynamics emerging from a large number of 
interacting agents have raised considerable interest 
\cite{Stigler}.  
The resulting phenomena often are difficult 
to capture in simple models, in particular equilibrium 
concepts fail in describing specific dynamical properties
of markets. Among such phenomena are expectation bubbles 
and subsequent crashes of expectation driven quantities
\cite{Markowitz,Kirman,Sornette}.  

Recently, fruitful attempts have been made to understand 
such emergent phenomena in systems of many interacting agents,
e.g. \cite{Solomon,Huberman,ContBouchaud}. 
Remarkably simple models can in principle capture essential 
features and, vice versa, provide toy systems that are  
as such interesting for statistical physics (but would 
hardly be written down from a pure physics perspective). 
One example of a simple model is the El Farol bar problem
\cite{Arthur}, most commonly known in a simplified formulation 
as the minority game \cite{minority}. In this game, agents 
split into two groups, however, only being in the minority 
group is rewarded and thus beneficial for each agent. 
This leads to a globally frustrated state as 
every single agent will try to reach this state. 
Similar interactions occur in real markets as, for example, 
it is often desirable to be in the minority when buying or 
selling a certain commodity. The minority game describes the 
evolution of strategic choices in a collective system within
such a simple frustrated boundary condition. Beyond this 
aspect, the elementary minority game does not provide a 
model for the more detailed dynamics of financial markets,
e.g. the dynamics of prices, as it is a very simplistic model. 
A recent attempt to include more general aspects of market 
dynamics into the minority game shows that quite complicated 
extensions are necessary, at the price of dropping the 
elegant simplicity of the original minority game \cite{Zhang}.

A complementary class of dynamical models for markets 
is formed by recent models of stock markets where 
trading agents are simulated, including an explicit price
formation process, e.g.\ \cite{ContBouchaud,Palmer,LuxNature}.     
A particularly realistic model is the one by Lux and Marchesi
\cite{LuxNature,Lux} who classify the agents into
two basic strategies (``fundamentalists'' and 
``optimistic/pessimistic chartists''). Prices form 
as a result of the trading activity, and strategies 
are chosen by each agent on the basis of profitability 
w.r.t.\ past actions. The resulting model reproduces 
several non-trivial properties observed in real markets, 
as power-law distributed returns $ret(t) = 
\ln(p(t)) - \ln(p(t-1))$ of prices $p(t)$, 
and a high autocorrelation 
of price volatility. Also, on the level of strategies, 
it shows the phenomenon that the number of chartists 
(or noise traders) correlates with phases of high volatility,
as also seen in real markets. The Lux-Marchesi-Model 
therefore is quite successful. From the theoretical point of 
view its only drawback is a high complexity with more than 10 
free parameters and considerable tuning. While this does not 
matter in economic applications of the model, a theoretical 
analysis of the basic mechanisms at work is not easy, facing 
a large number of details in the model.  

In this article, therefore, a third path will be sketched, 
formulating a model with maximum simplicity, while 
including details of strategic interactions in a market. 
It can be viewed as an extremely simplified version of a 
stock market model in the form of a spin model with only 
two alternatives, designed to simulate 
the dynamics of expectations in systems of many agents. 
While spin models have a tradition in economic 
theory \cite{Foellmer,Blume,Aoki,Kaizoji}, the critical 
properties of such models are often difficult to relate 
to real economic situations, at least without tuning. 
In particular, large fluctuations as often seen in real
economic data \cite{Gopikrishnan,Bouchaud} 
usually appear only near the critical point
of spin models. One possibility to capture these features in 
a spin model is to reformulate a stock market model in terms
of a generalized spin models as, for example, demonstrated  
by Chowdhury and Stauffer who recast the Cont-Bouchaud-Model
in terms of ``super spins'' \cite{Stauffer}. Also the 
random field Ising model can be related to critical properties 
of stock market dynamics \cite{Iori}.  

Let us here take a different approach and rather start from 
scratch by asking, what are the basic forces at work? 
There are at least two major conflicting forces seen in 
economic action: 
\noindent \\ 
1. ``Do what your neighbors do'', as often seen in the  
action of noise traders and as modeled in the herd behavior 
of strategic choices in the Lux-Marchesi-Model, and: 
\noindent \\ 
2. ``Do what the minority does'', as often followed by 
traders with knowledge about fundamental values and 
as modeled in the minority game.  

We here combine these two conflicting interactions in 
a simple spin model: On the one hand, the neighbor 
interaction is represented as a standard nearest 
neighbor interaction of a spin model. On the other hand,  
a coupling to the minority as a global observable 
is introduced by a coupling to the global magnetization
of the spin system. 

Consider a model with $i=1,...,N$ spins with orientations 
$S_i(t) = \pm 1$. The dynamics of the spins depends on the 
local field $h_i(t)$, for simplicity assume that each spin 
is updated with a heat-bath dynamics according to  
\begin{eqnarray} 
S_i(t+1) &=& +1 
\;\;\; \mbox{with} \;\;\; 
p = 1/\left[1+\exp\left(-2\beta h_i(t)\right)\right]
\nonumber \\ 
S_i(t+1) &=& -1 
\;\;\; \mbox{with} \;\;\; 
1-p. 
\end{eqnarray}  
The local field containing the interactions discussed above
is further specified by  
\begin{equation} 
h_i(t) = \sum_{j=1}^NJ_{ij}\;S_{j} - 
\alpha \; C_i(t) \; \frac{1}{N} \sum_{j=1}^{N} S_j(t). 
\end{equation}  
The first term is chosen as a local Ising Hamiltonian
with nearest neighbor interactions $J_{ij}=J$ and  
$J_{ij}=0$ for all other pairs. This term thus induces 
local ferromagnetic order. The second term is a global 
coupling to the magnetization of the system with a 
coupling $\alpha>0$. Its sign determines the strategy of 
agent $i$. It is specified separately by a second spin 
$C_i(t)$, representing the strategy of agent $i$ with 
respect to the magnetization. In particular, this second
coupling allows for the case of spins frustrated across
scales, seeking ferromagnetic order locally, but  
anti-ferromagnetic order globally.  
Similar couplings to the total magnetization are known for 
magnets with dipole interactions as ``demagnetizing field''     
which in the context of market models are reminiscent of 
the welfare state mode of the Levy-Solomon-Huang-Model 
\cite{Huang}. 

What is the basic dynamics of this system? 
Several scenarios can be realized depending on value and 
dynamics of the strategy spin field. Let us consider the 
simplest possibilities. First, consider the case where 
the strategy is always $C_i(t)=1, \;\;\; \forall i,t$. 
Each agent then, besides the local ferromagnetic 
coupling to the neighbors, has an anti-ferromagnetic 
coupling to the magnetization. This, in a sense  
follows a motivation reminiscent of the minority game, 
inducing a force to align with the minority of spins in 
the system. In contrast to the minority game, however, 
it is complemented by a conflicting ferromagnetic 
term which the minority game lacks.  
The dynamics with $C_i(t)=1$ then corresponds to 
traders, who, in addition to a basic level of 
ferromagnetic noise trading, have a desire to join 
the global minority, for example in order to invest in 
possible future gains. Thus the $C_i(t)=1$ traders 
can be called fundamentalists. If every agent follows 
this strategy, the global dynamics of the model will 
quickly lead to a near-vanishing magnetization, 
even for lower-than-critical temperatures $T<T_c$,   
resulting in a ``soft'' conserved order parameter  
Ising model. Large fluctuations in the magnetization 
are suppressed in this case.   

The dynamics of the system becomes more interesting once 
agents are allowed to follow two different 
strategies s.t.\ also the strategy $C_i(t)=-1$ occurs in 
the spin system which corresponds to a ferromagnetic   
coupling to the global magnetization. This strategy is 
called the chartist strategy as agents tend to follow 
the majority of traders. Initializing the model with a 
fixed ratio of the two strategies, one observes a transition 
between the conserved order parameter regime, where 
fundamentalists dominate, and a globally magnetized state 
with strongly attractive fixed points at the two points of 
maximum magnetization. This is where chartists dominate.     

It remains to define the transition rule between the 
two strategies as each trader will tend to choose an 
optimal strategy which in general will not result in  
a fixed ratio of the two strategies. Let us
consider the simplest possible scenario for local 
strategy changes. An agent who is in the majority group 
will often tend to switch to the minority group, 
e.g.\ in order to opt for future prospective 
returns of that not-yet fashionable commodity
(and possibly escape a future crash of its currently
popular good). As such an option is only available 
at a cost (the effort to obtain information, 
the potential risk as it affects prospective future return, etc.), 
the corresponding term will increase the local energy of the agent. 
Vice versa, let us assume that an agent who is in the minority 
(thus expecting future returns) might be unsatisfied with 
present returns, the more so, the larger the majority group 
of agents is, whose returns correlate with the current popularity 
of the commodity. Again, this option comes at a price accounted 
for by a positive energy term. The scenario is summarized 
under the simple assumption that a trader in the majority  
will always choose strategy $C_i(t)=1$, while a minority 
trader will always choose $C_i(t)=-1$: Each trader chooses 
the risky strategy for the prospect of higher returns.   
The dynamics of the strategy spins then is 
\begin{eqnarray} 
C_i(t+1) = -C_i(t) 
\;\;\; \mbox{if} \;\;\; 
\alpha \; S_i(t) \; C_i(t) \; \sum_{j=1}^{N} S_j(t)<0.
\end{eqnarray}  
Each agent thus faces a certain penalty for his strategic 
action, which increases with the absolute value of 
magnetization. This is motivated by the fact that 
in general speculation costs rise in times with high 
volatility. 

A particularly simple model is obtained, if the strategy 
adjustment is assumed to be done instantaneously.  
Then, the strategy spin drops out altogether and we 
obtain a simple spin model with the local field defined as 
\begin{equation} 
h_i = \sum_{j=1}^NJ_{ij} \; S_j 
- \alpha \; S_i \; \left| \frac{1}{N} \sum_{j=1}^{N} S_j \right|  
\end{equation}  
with a global coupling constant $\alpha>0$. 
While the first term tends to align the spin with its neighbors, 
the second term tends to encourage a spin flip when magnetization 
gets large.  

The dynamics of this model is characterized by 
metastable phases of approximate undercritical Ising dynamics 
and intermediate phases of rapid rearrangements, 
reminiscent of overcritical dynamics. An example of 
subsequent snapshots in a $2d$ system is shown in Fig.\ 1. 
\begin{figure}[htb]
\let\picnaturalsize=N
\def\picsize{85mm}
\def\picfilename{bornholdt1.eps}
\ifx\nopictures Y\else{\ifx\epsfloaded Y\else\input epsf \fi
\let\epsfloaded=Y
\centerline{\ifx\picnaturalsize N\epsfxsize \picsize\fi
\epsfbox{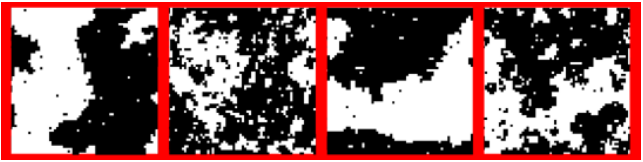}}}\fi
\medskip 
\caption{
Subsequent snapshots of the spin dynamics of a 
32x32 lattice at undercritical temperature, taken 
at $t=611,\; 2046,\; 2913,\; 3527$ sweeps (Monte Carlo updates).   
The first and third snapshots are taken during a metastable 
phase, the second and fourth are in the turbulent regime, 
where memory of global spatial structures disintegrates.  
}   
\end{figure}
Here, a 32x32 lattice of the general version of the model 
as defined in eq.\ (2) and where the strategy spins are updated 
according to eq.\ (3) is shown. It is simulated at temperature 
$T=1/\beta=1.5$ and with couplings $J=1$ and $\alpha=4$, 
using random serial and asynchronous heat bath updates
of single sites. Note that the temperature is below the 
critical temperature $T<T_c=2.269$ of the case $\alpha=0$.  
For every spin $S_i$ that is updated, 
the strategy spin $C_i$ is updated subsequently.  
The way the change of strategies is defined here, 
the model is to a large extent identical to the simplified 
version eq.\ (4), but in addition keeps track of 
the ratio of strategy choices.  
A metastable phase occurs at randomly frozen finite 
magnetization values, which, in the language of market 
dynamics, is the analogue of a bubble-related value
of a good that emerged without any fundamental cause.  
Identifying a spin $S_i=+1$ with a buyer and $S_i=-1$ 
with a seller, the spin value can be viewed as the demand of 
an agent $i$ and the total excess demand, or magnetization, 
can be related to global price changes \cite{ContBouchaud}.  
Thus interpreting magnetization $M(t)=\frac{1}{N} \sum_{j=1}^{N} S_j$ 
as a measure of price, it is interesting to plot its logarithmic 
relative change $ret(t) = \ln(M(t)) - \ln(M(t-1))$, 
where intermittent phases are nicely visible (Fig.\ 2). 
\begin{figure}[htb]
\let\picnaturalsize=N
\def\picsize{85mm}
\def\picfilename{bornholdt2.eps}
\ifx\nopictures Y\else{\ifx\epsfloaded Y\else\input epsf \fi
\let\epsfloaded=Y
\centerline{\ifx\picnaturalsize N\epsfxsize \picsize\fi
\epsfbox{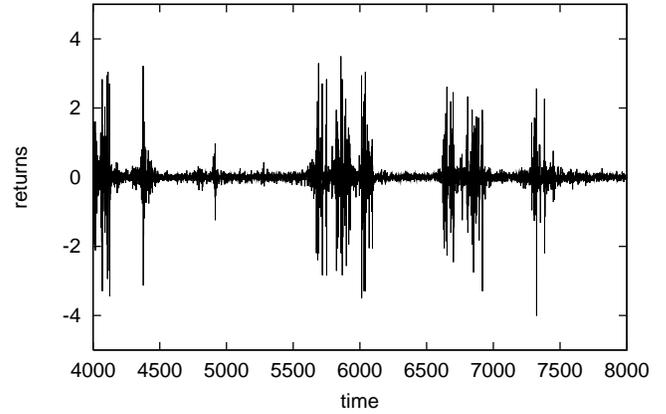}}}\fi
\caption{Return or logarithmic relative change of 
magnetization of the $2d$ model as defined above.}
\end{figure}
Comparing this observable to economic returns, features
as non-Gaussian fluctuations and clustered phases of large 
fluctuations are observed here that are also known from 
real economic data. More specifically, 
the cumulative distribution of returns 
as derived from the magnetization of this model shows a 
pronounced power-law scaling (Fig.\ 3). 
\begin{figure}[htb]
\let\picnaturalsize=N
\def\picsize{85mm}
\def\picfilename{bornholdt3.eps}
\ifx\nopictures Y\else{\ifx\epsfloaded Y\else\input epsf \fi
\let\epsfloaded=Y
\centerline{\ifx\picnaturalsize N\epsfxsize \picsize\fi
\epsfbox{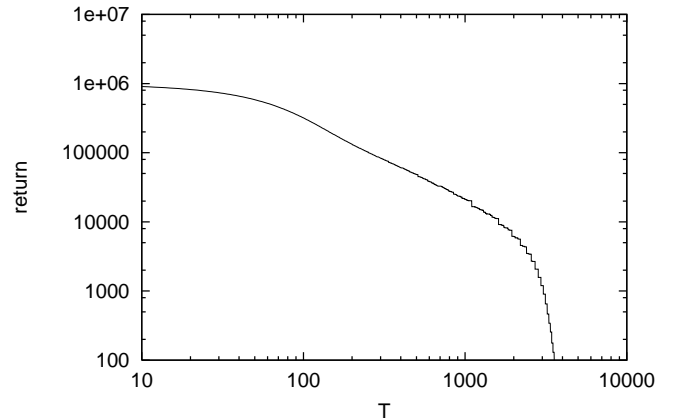}
}}\fi
\caption{Cumulative distribution of absolute returns exhibits 
power-law scaling. The 32x32 model is defined as above with parameters 
$T=1.0$ and $\alpha=8$ and sampled over $10^6$ sweeps.}   
\end{figure}
The second feature, also seen in Fig.\ 2, are phases of 
high volatility that are strongly clustered. To quantify this, 
the corresponding autocorrelation of absolute returns is shown 
in Fig.\ 4.  
\begin{figure}[htb]
\let\picnaturalsize=N
\def\picsize{85mm}
\def\picfilename{bornholdt4.eps}
\ifx\nopictures Y\else{\ifx\epsfloaded Y\else\input epsf \fi
\let\epsfloaded=Y
\centerline{\ifx\picnaturalsize N\epsfxsize \picsize\fi
\epsfbox{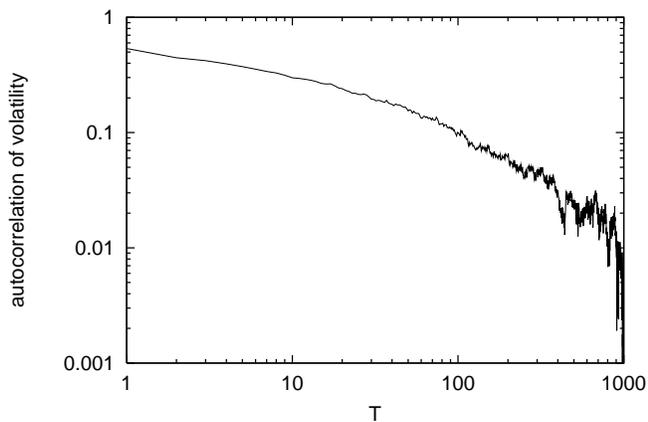}}}\fi
\caption{Autocorrelation of absolute returns as a measure of 
volatility clustering. Model parameters are as in previous figure.}   
\end{figure}
An interesting numerical experiment is to look at the 
first formulation of the above model, where strategy changes 
are explicitly tracked in the strategy spin variable $C_i(t)$: 
Each agent switches its strategy to the opposite value, if 
the new strategy will be more risky (results in higher energy). 
This allows to see how strategy choices correlate to the 
phases of high volatility. Indeed, phases of high variance 
in changes of magnetization correlate with the number of agents 
that play the chartist strategy (Fig.\ 5). 
\begin{figure}[htb]
\let\picnaturalsize=N
\def\picsize{85mm}
\def\picfilename{bornholdt5.eps}
\ifx\nopictures Y\else{\ifx\epsfloaded Y\else\input epsf \fi
\let\epsfloaded=Y
\centerline{\ifx\picnaturalsize N\epsfxsize \picsize\fi
\epsfbox{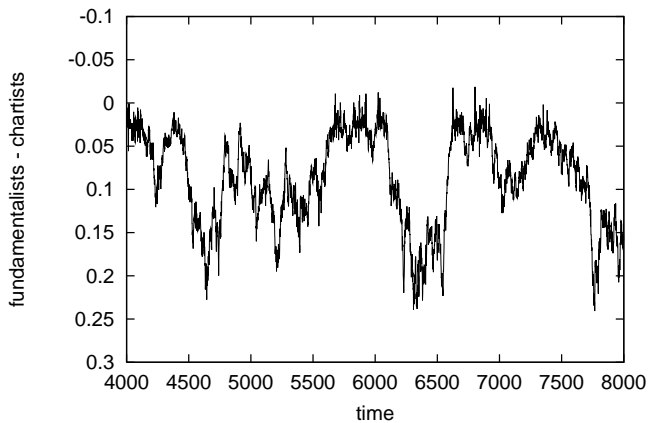}}}\fi
\caption{Fraction of chartists in the same run as shown in 
Fig.\ 1. A high number of chartists coincides with phases of 
high volatility.}   
\end{figure}

The simple spin model studied here reproduces major 
observational features of real economic markets. Due to its 
simplicity, it offers a chance to understand basic 
mechanisms at work in economic systems of many agents
competing for commodities, as well as the possibility of 
analytic tractability that is often not given in complex agent 
simulations of the same phenomena. 

We here considered the model in $2d$ which is the 
lowest non-trivial number of dimensions. 
Many simple variants of the model are an obvious target 
of future research: Other dimensions or neighborhoods, 
the spin glass variants with random couplings, 
and other strategy changing rules (when including an 
explicit strategy spin), as well as a general $q$-state model 
with more than 2 states.  

The spin model studied here shows intermittent 
behavior as a result of competition between conflicting 
local and global interactions. While exhibiting 
a phase transition at higher temperatures, it 
is considered here in the undercritical phase. As a result,
the basic observation occurs in a wide temperature range and
does not depend on tuning: Self-organization 
of the spin system to a regime dominated by metastable 
phases with intermittent disorder.

\end{document}